# MATHEMATICAL MODEL OF AN INDUCTIVE MEASURING CELL FOR CONTACTLESS CONDUCTOMETRY[*]

by Yury S. Semenov,

Moscow Institute of Physics and Technology (State University)

PREFACE. First of all, I would like to express my deep gratitude to those who are reading this text. Then, I must apologize and have to warn you: presented text is a complete but not final version of translation, it still requires editing. English is not my native language, so I will be very grateful for your comments and corrections. Please, feel free to email me (semenov.yury@gmail.com).

I hope the model will be useful for you. If you have any questions, do not hesitate to send me a mail. Have a great read!

*Yury S. Semenov*

ABSTRACT. A research of inductive conductometric cell is presented. An equivalent circuit and a mathematical model of inductive cell are given in the article. The model takes into account sample-coil capacity (i.e. capacity formed by the coil and the sample under study) and eddy currents. It is sample-coil capacity that makes inductive cell applicable for measurement of electrical conductivity of low conductive samples (specific conductance is less than $1\,S/m$). The model can be used to calculate impedance of inductive cell for different characteristics of sample, materials and dimensions of cell without numerical solving of partial differential equations. Results of numerical simulation were verified by experiment for several devices with inductive cell. Some features that an engineer has to hold in mind while designing a conductometer based on inductive cell are discussed. Presented model can be useful for those who study inductively coupled plasma.

INTRODUCTION

It is often necessary to monitor concentration of electrolytes in a solution, amount of contaminants in samples or components of a mixture in real time. A very convenient method to solve this problem is

---

measuring of the specific conductance of a sample (conductometry). However, an application domain of traditional contact conductometers is limited by corrosion of electrodes and near-electrode processes. One can overcome the limitation by using contactless transducers. There are three main types of transducers (transformer, capacitive and inductive cell), others are combinations of them. All types of contactless transducers operate on alternating current. If the sample is aqueous solution, one should use frequencies from about 10 Hz to 10 MHz. Lower frequencies is inappropriate due to decrease in sensitivity, higher – due to increase in dielectric loss and Debye–Falkenhagen effect. This article deals with an inductive cell.

Inductive cell is an inductive coil wound round a dielectric tube. The sample under study is placed inside the tube. It is believed that such transducers can be effective when operating with a high-conductive samples (metals) only. In this case, the coil impedance changes due to eddy currents induced in the sample. This mode is well researched. There are formulas that relate impedance of the inductive cell with the conductivity of the sample [1, 2]. But in case of low-conductive samples, such as aqueous solutions of various electrolytes, eddy currents are not high enough to change the impedance of the cell significantly. That is why inductive cells are not commonly used to operate with solutions.

However, experimental results were out of accord with theoretical conclusion stating that inductive cells don not work in case of solutions. It was shown that such type of transducers have a sensitivity increase at frequencies about 1 MHz. It could not be described in terms of eddy current. Later it had been shown that inductive cells have a remarkable feature: there is the parasitic capacitance between the coil and the sample under study. It is parasitic capacity that results in appearance of the increase [2].

Usually engineers try to eliminate effects of the capacity by using electromagnetic shielding despite of decrease in sensitivity (shield, as a conductor, may significantly affect properties of a transducer). This happens because there is no mathematical model that includes coil-sample capacity and simultaneously allows to calculate relationship between impedance of cell and properties of sample before the cell produced. Book written by prof. B.A. Lopatin [2] (and English version of it [3]) describes various equivalent circuits of inductive cell, some of them takes into account coil-sample capacity. However, it is impossible to calculate parameters of circuits, one have to measure them for produced transducer.

The aim of this work is to develop an equivalent circuit and mathematical model of inductive cell that will describe the effect of coil-sample capacity on impedance of the transducer and verify theoretical conclusions in experiments. The model should be as simple as possible and allow to calculate impedance of an inductive cell in terms of electrical properties of materials and a sample and dimensions of the cell without numerical solving of partial differential equations. It should be possible to analyze the model analytically.

# THEORY

## AN EQUIVALENT CIRCUIT

Let us consider the processes occurring inside an inductive cell, if there is an isotropic conductive sample placed into the cell. Coil is assumed long enough to neglect edge effects.

The electric field inside the coil is originated not by the presence of an alternating magnetic field only. Assume that the frequency of the voltage source connected to the cell is zero. There is distribution of potential along turns of the coil. Each new turn has potential greater than the potential of the previous one. Therefore, electric field exists not only outside but inside the coil. At every point inside the cell one may consider an electric strength vector as a superposition of radial and axial components due to cylindrical symmetry of the cell. Radial component charges coil-sample capacitor formed by material of tube wall, turns of the coil and lateral surface of the sample. Axial component produce an axial electric current charging a capacitor formed by ends of the sample. In case of an alternating current the situation is the same but eddy electric field appears due to axial magnetic field. Charging of edge and coil-sample capacitors affects the electric field distribution and, therefore, impedance of the cell as well as eddy electric field affects axial magnetic field.

I assume the coil as a system with lumped parameters (Fig. 1 and Fig. 2a). I consider a single-layer long straight coil wound turn to turn with the lowest possible tilt of turns. Thickness of measuring cell wall is much greater than diameter of the wire, and inner radius of the cell is much greater than thickness of the wall. As an approximation, I assume that the potential is constant along a single turn and changes from turn to turn stepwise, turns have a zero slope, i.e. every turn is a ring. And as I consider long coil, I assume that there is no edge effects, so there is no distribution in values of parameters of the circuit along the cell. Within the bounds of the approximation I suggest an equivalent circuit of the transducer presented at Fig. 2.

Capacitance $C_1$ is a capacitance between two nearby turns. $C_2$ represents capacitance of an element of the coil-sample capacitor. Number of capacitors $C_2$ is equal to the number of turns of the coil $N$. Inductance $L_1$ is an inductance of the coil divided by the number of elementary units shown at Fig. 2c as well as resistance $R_1$ (resistance of the wire). $R_S$ is a parameter describing electrical resistivity of a sample. Impedance $Z_1$ is formed by $L_1$ and $R_1$ connected in series and $C_1$ in parallel to them, $Z_2$ is just the impedance of $R_S$, $Z_3$ is impedance of $C_2$. It is easy to calculate that the number of elementary units is equal to $(N-1)$. So $L_1$ is approximate inductance of a turn of the coil as well as $R_1$ is the resistance of a turn. $C_S$ is capacitor formed by ends of the sample, $Z_{cs}$ is impedance of $C_S$. And the last element is capacitor $C_W$ representing parasitic capacitance of wire that connects the coil to circuits of an impedance meter. $C_W$ is absent at Fig 2b.

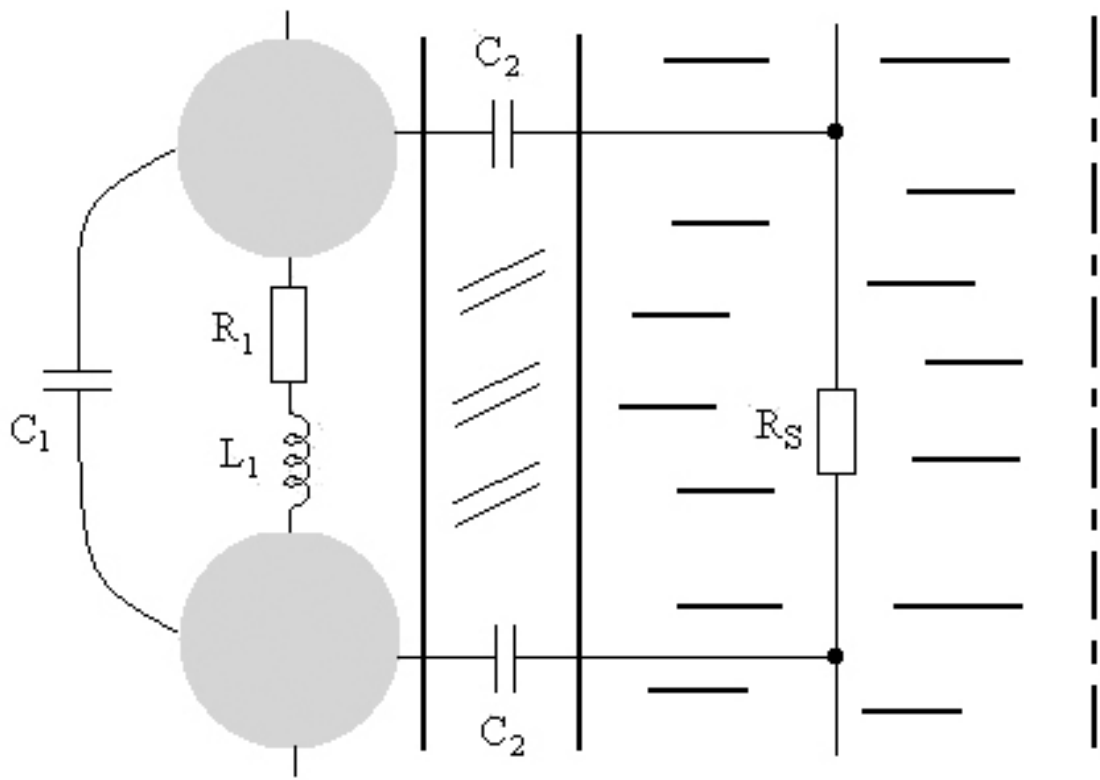

Fig. 1. Diagram that illustrates the physical meaning of the equivalent parameters (figure not to scale). Two nearby turns are shown (explanation in the text).

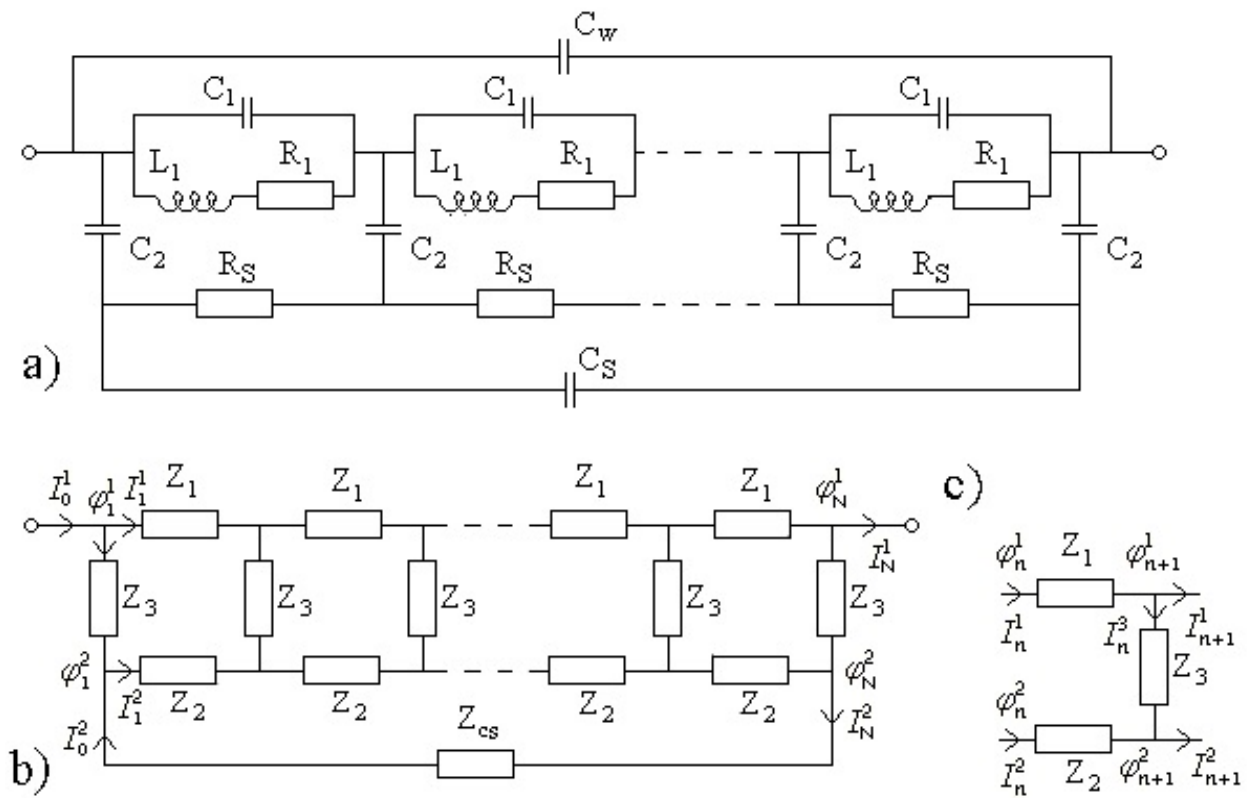

Fig. 2. The equivalent circuit of the inductive measuring cell (explanation in the text): a) complete variant of the equivalent circuit, b) circuit used in calculating of impedance of the cell with a sample, c) – elementary unit.

## PARAMETERS ESTIMATING

To estimate parameters of the equivalent circuit quantitatively I use following approach. A reader may use another one. All formulas are given in SI units. I consider situation when lengths of the coil and the sample are equal, if they are not, one should use the shortest one as a length of the coil and the sample (and do not forget to correct the number of turns respectively!). I suppose that there is no gasp between lateral surface of the sample and inner surface of the cell. If there is gasp, one should take it into account by decreasing capacitance of the coil-sample capacitor.

Capacitance $C_1$ does not depend on the sample, it formed by two nearby turns (rings, in terms of the equivalent circuit). So assuming that two nearby turns are two straight parallel wires of same diameters and as long as the circumference of a turn is the simplest way to estimate $C_1$. The assumption is reasonable, because radius of the turn much greater than diameter of the coil wire. I do not consider dielectric properties of isolation of the wire or material of cell wall, because each of them in majority of cases has relative permittivity less than 10.

$$C_1 = 2\pi^2 R \varepsilon_2 \varepsilon_0 \bigg/ \ln\left(\frac{1}{1-\sqrt{(1+\delta/d)^2-1}}\right),$$

here $\varepsilon_2$ is relative permittivity of the medium round the turns (usually it is air), $\delta$ – shortest distance between surfaces of the metal core of the turns (doubled thickness of wire isolation), $d$ - diameter of the metal core of the wire, $R$ - inner radius of the cell (approximately equals to the radius of the turns due to thin cell wall).

The approximation is very rough, but at frequencies lower than 10 MHz it usually does not affect the accuracy of coil impedance calculation. Anyway, one may just measure this parameter by placing two rings made from wire, which is same to coil wire, at the distance, that equals to the distance between turns, round the tube. Capacitance between the rings is $C_1$.

If $\delta \ll d$ there is simpler formula: $C_1 \approx 2\pi^2 R \varepsilon_2 \varepsilon_0 \sqrt{\dfrac{d}{2\delta}}$.

The coil-sample capacitor is formed by lateral surface of the sample and inner surface of the cell, so it is just cylindrical capacitor. It is represented by $N$ parallel capacitors $C_2$, so

$$C_2 = \frac{2\pi\varepsilon_0\varepsilon_1 l}{\ln(1+h/R)} \cdot \frac{1}{N} = \frac{2\pi\varepsilon_0\varepsilon_1 d}{\ln(1+h/R)},$$

here $\varepsilon_1$ is relative permittivity of the material of the cell wall, $h$ - thickness of the wall, $l$ - length of the coil. As $h \ll R$, then $C_2 \approx 2\pi\varepsilon_0\varepsilon_1 R d/h$.

One may measure $C_2$ by pressing metal foil to the inner surface of the cell and connecting the foil to one lead of a capacitance meter. Other lead of capacitance meter should be connected to leads of the coil. Readings of the meter should be divided by $N$.

As mentioned above, $L_1$ and $R_1$ are inductance ($L_0$) and resistance ($R_0$) of the empty coil divided by $(N-1)$.

I use standard formulas to estimate $L_0$ and $R_0$. I do not take into account skin and proximity effects in the wire. These effects impact the impedance of the cell in case of resonance only. If the resonance occurs, there are strong requirements for accuracy of estimating (or measuring) other model parameters too.

$$R_1 = \frac{R_0}{N-1}, \quad L_1 = \frac{L_0}{N-1}$$

Of course, one may just measure $L_0$ and $R_0$ at different frequencies or use litz wire.

If there is a sample inside the cell, one has to take into account skin effect into the sample. It affects the impedance of the cell dramatically, especially, in case of high-conductive samples. So it is necessary to calculate correction to inductance of the cell.

Here I consider inductance of the cell with the sample inside. At first, I determine distribution of a magnetic field in the sample (analogous problem for electric field is solved in a book [4]).

Set of Maxwell's equations and constitutive relations (including Ohm's law) can be reduced to an equation in magnetic field $H$. As the cell has axial symmetry, then in cylindrical coordinates (and complex notation) the equation is:

$$\frac{\partial^2 H}{\partial r^2} + \frac{1}{r}\frac{\partial H}{\partial r} + (-i\omega\sigma\mu\mu_0 - \varepsilon_S\varepsilon_0\mu\mu_0(i\omega)^2)H = 0,$$

where $i^2 = -1$, $\mu$ is relative permeability of the sample, $\varepsilon_S$ - relative permittivity of the sample, $\sigma$ - specific conductance of the sample, $\omega$ - angular frequency of a current flowing through turns of the coil.

The equation can be reduced to Bessel's differential equation of zero order by a substitution of variables $x = r\sqrt{A}$, where $A = -i\omega\sigma\mu\mu_0 - \varepsilon_S\varepsilon_0\mu\mu_0(i\omega)^2$.

General solution is:

$$H(r\sqrt{A}) = bJ_0(r\sqrt{A}) + cY_0(r\sqrt{A}).$$

Constant $c$ is zero due to finite value of $H$ at the axis of the cell.

Constant $b$ can be find with the help of boundary condition: magnetic field inside the cell but out of the sample is equal to the field of ideal empty coil:

$$b = \frac{NI_0}{lJ_0(R\sqrt{A})},$$

where $N$ - number of turns of the coil, $l$ - length of the coil, $R$ - inner radius of the coil, $I = I_0 e^{i\omega t}$ - current flowing through turns of the coil.

Therefore, $H(r,t) = \frac{NI_0 J_0(r\sqrt{A}) e^{i\omega t}}{lJ_0(R\sqrt{A})}$.

The next step is to calculate instantaneous magnetic flux through turns of the coil $\Phi$ and then inductance of the coil with the sample $L$:

$$\Phi(t) = N\int_0^R 2\pi r \mu\mu_0 H(r,t) dr = \frac{2\pi N^2 I_0 \mu\mu_0 R J_1(R\sqrt{A}) e^{i\omega t}}{l\sqrt{A} J_0(R\sqrt{A})},$$

$$L = \frac{\Phi(t)}{I(t)} = \frac{2\pi N^2 \mu\mu_0 R J_1(R\sqrt{A})}{l\sqrt{A} J_0(R\sqrt{A})}.$$

Finally, $L_1 = \frac{L_0}{(N-1)} \frac{2 J_1(R\sqrt{A})}{R\sqrt{A} J_0(R\sqrt{A})}$.

The last relation is proper for high- and low-conductive samples. The inductance written in that form describes changes in flux due to eddy currents (corrections to the real part) and additional ohmic losses into the sample (imaginary part).

In case of low-conductive non-magnetic samples (i.e. $|R\sqrt{A}| \ll 1$) the Bessel's function can be expanded into a series to second order:

$$L = \frac{2\pi N^2 \mu\mu_0 R J_1(R\sqrt{A})}{l\sqrt{A} J_0(R\sqrt{A})} = \frac{\pi N^2 \mu\mu_0 R^2 \cdot 2 J_1(R\sqrt{A})}{l \cdot R\sqrt{A} J_0(R\sqrt{A})} = L_0 \frac{2 J_1(R\sqrt{A})}{R\sqrt{A} J_0(R\sqrt{A})} \approx L_0 \left(1 - \frac{i\omega R^2 \sigma\mu\mu_0}{8}\right).$$

And then, concerning the impedance of the cell (without coil-sample capacitance), I have:

$$Z = R_0 + i\omega L = R_0 + i\omega L_0 + \frac{R^2 \omega^2 \sigma\mu\mu_0 L_0}{8}.$$

To estimate $R_s$ one has to know disturbance of the axial component of the electric field along the radius of the coil.

I find it analogous to considering inductance of the coil. Here I skip intermediate calculations that are the same to written above. It is important that there is no volume charge in the bulk under consideration.

$$E(r) = \frac{U}{d} \cdot \frac{J_0(r\sqrt{A})}{J_0(R\sqrt{A})},$$

where $U$ is turn-to-turn voltage that "measured" at lateral surface of the sample.

So total axial current is easy to find by integrating over radius: $\tilde{I} = \frac{U\pi R^2 \sigma \cdot 2J_1(R\sqrt{A})}{d \cdot R\sqrt{A} J_0(R\sqrt{A})}$

Finally,

$$R_S = \frac{U}{\tilde{I}} = \frac{d}{\pi R^2} \frac{R\sqrt{A} J_0(R\sqrt{A})}{2J_1(R\sqrt{A})} \frac{1}{\sigma}.$$

If $|R\sqrt{A}| \ll 1$, then it is possible to use an approximation $R_S \approx \frac{d}{\pi R^2} \frac{1}{\sigma} \left(1 + \frac{i\omega\mu_0 \sigma R^2}{8}\right)$ or go even further and just use Pouillet's law $R_S \approx \frac{d}{\pi R^2} \frac{1}{\sigma}$.

I estimate capacitance formed by ends of the sample ($C_S$) as a capacitance of a parallel-plate capacitor:

$$C_S = \varepsilon_S \varepsilon_0 \pi R^2 / l.$$

I do not take into account skin effect or edge effects. It is rough approximation but as well as turn-to-turn capacitance ($C_1$) approximation it usually does not affect the accuracy of coil impedance calculation.

To sum up, parameters of the equivalent circuit are given below:

$$C_S = \frac{\varepsilon_S \varepsilon_0 \pi R^2}{l}, \qquad R_S = \frac{d}{\pi R^2} \frac{R\sqrt{A} J_0(R\sqrt{A})}{2J_1(R\sqrt{A})} \frac{1}{\sigma}, \qquad C_1 \approx 2\pi^2 R \varepsilon_2 \varepsilon_0 \sqrt{\frac{d}{2\delta}},$$

$$C_2 \approx 2\pi \varepsilon_0 \varepsilon_1 \frac{Rd}{h}, \qquad L_1 = \frac{L_0}{(N-1)} \frac{2J_1(R\sqrt{A})}{R\sqrt{A} J_0(R\sqrt{A})}, \qquad R_1 = \frac{R_0}{N-1},$$

$$A = -i\omega\sigma\mu\mu_0 - \varepsilon_S \varepsilon_0 \mu\mu_0 (i\omega)^2.$$

In case of non-magnetic aqueous solutions at frequencies lower than 10 MHz and $R < 1\,cm$ it is possible to use an approximation $\frac{R\sqrt{A}J_0(R\sqrt{A})}{2J_1(R\sqrt{A})} \approx \frac{2J_1(R\sqrt{A})}{R\sqrt{A}J_0(R\sqrt{A})} \approx 1$. In other words, it is possible to neglect skin effect and eddy currents in the sample.

Regarding $C_W$ it should be noted that it is almost impossible to estimate or measure it. Even in case of success this parameter usually drifts in a wide range of values. So the best way is to minimize $C_W$ by construction of a meter and the cell as far as possible, it will diminish influence of $C_W$ on the impedance of the cell. Also it is possible to compensate $C_W$ by connecting auxiliary capacitor to leads of the cell in parallel.

SOLVING PROCEDURE

Below I use complex notation for all variables. Total cell impedance $Z$ (without parasitic capacitance of the wire $C_W$, Fig 2b) can easily be calculated with the help of Kirchhoff's circuit laws applied to the elementary unit shown at Fig. 2c.

$$\begin{cases} \varphi_n^1 - \varphi_{n+1}^1 = Z_1 I_n^1 \\ \varphi_n^2 - \varphi_{n+1}^2 = Z_2 I_n^2 \\ \varphi_{n+1}^1 - \varphi_{n+1}^2 = Z_3 I_n^3 \\ I_n^1 = I_n^3 + I_{n+1}^1 \\ I_n^2 + I_n^3 = I_{n+1}^2 \end{cases}$$

Here upper index represents branch of the circuit, lower index is a number of a node. Potentials $\varphi_n^m$ and currents $I_n^m$ are presented at Fig. 2b, 2c.

The set of equations can be reduced to a recurrent equation in potential of the first or the second branch. And after index shift the equation is:

$$\varphi_{n+4}^2 - \varphi_{n+3}^2 \cdot \left(4 + \frac{Z_2}{Z_3} + \frac{Z_1}{Z_3}\right) + \varphi_{n+2}^2 \cdot 2\left(3 + \frac{Z_2}{Z_3} + \frac{Z_1}{Z_3}\right) - \varphi_{n+1}^2 \cdot \left(4 + \frac{Z_2}{Z_3} + \frac{Z_1}{Z_3}\right) + \varphi_n^2 = 0$$

To find boundary conditions it is necessary to take into account $Z_{cs}$ and the first (leftmost) impedance $Z_3$ that are not a part of elementary unit (Fig. 2c).

$$\begin{cases} I_0^1 = I_1^1 + \dfrac{\varphi_1^1 - \varphi_1^2}{Z_3} \\ I_0^2 = I_1^2 - \dfrac{\varphi_1^1 - \varphi_1^2}{Z_3} \end{cases}$$

Boundary conditions are:

$$\begin{cases} I_0^2 = I_N^2 \\ I_0^1 = I \\ \varphi_N^2 - \varphi_1^2 = I_0^2 Z_{cs} \\ \varphi_1^1 - \varphi_N^1 = U \end{cases},$$

here $I$ is a current flowing through the coil, $U$ is potential difference between leads of the coil, i.e. $Z = \dfrac{U}{I}$. It is worth to note that as currents satisfy the relation $I_0^2 = I_N^2$, then the equation $I_0^1 = I_N^1$ is valid automatically.

After solving the recurrent equation (procedure is described in book [5]) and performing some rearrangements I derived a formula for the cell impedance (without parasitic capacitance of the wire):

$$Z = \dfrac{\alpha\beta + \chi\left(\dfrac{\left(\lambda_2^{N-1} - \lambda_1^{N-1} + \lambda_2 - \lambda_1\right)}{\left(\lambda_2^N - \lambda_1^N\right)} - 1\right)}{\alpha\gamma + \dfrac{2}{Z_1}\left(\dfrac{\left(\lambda_2^{N-1} - \lambda_1^{N-1} + \lambda_2 - \lambda_1\right)}{\left(\lambda_2^N - \lambda_1^N\right)} - 1\right)},$$

here I use following designations:

$$\alpha = \dfrac{Z_1 + Z_2}{Z_3}, \quad \beta = -(N-1)\dfrac{Z_{cs}}{Z_2}, \quad \chi = \dfrac{2\left[Z_{cs} Z_1 + (N-1)Z_2(Z_1 + Z_2)\right]}{Z_2^2},$$

$$\gamma = -\dfrac{1}{Z_1}\left(\dfrac{Z_1}{Z_2}\left(N - 1 + \dfrac{Z_{cs}}{Z_2}\right) + \dfrac{Z_{cs}}{Z_2}\right), \quad \lambda_{1,2} = 1 + \dfrac{\alpha}{2}\left(1 \pm \sqrt{1 + \dfrac{4}{\alpha}}\right).$$

Formulas are correct only if $\alpha \neq 0$ and $N \geq 2$, there are no another limitations. As I operate with general form of impedances notation, a reader may use these formulas to calculate impedance of a long circuit with arbitrary array of elements and, therefore, arbitrary $Z_1$, $Z_2$, $Z_3$, $Z_{cs}$. For the model I use following notation:

$$Z_1 = \dfrac{R_1 + i\omega L_1}{(R_1 + i\omega L_1)i\omega C_1 + 1}, \quad Z_2 = R_S, \quad Z_3 = \dfrac{1}{i\omega C_2}, \quad Z_{cs} = \dfrac{1}{i\omega C_S}.$$

Another forms of formula describing the cell impedance are possible, all forms are equivalent. In my opinion, given form is more convenient for theoretical analysis. Also I use it for computation.

Finally, the impedance of the transducer (including parasitic capacitance of the wire $C_W$) is given by $Z_{tr} = \dfrac{Z}{1 + i\omega C_W Z}$.

Sometimes it is necessary to take into account processes involving shielding of electric field by free charges of a solution and diffusion of charges. In that case one should add a concentration-dependent capacitance to $R_S$. Value of the capacitance can be estimated with the help of known electrochemical methods.

RESULTS OF CALCULATIONS AND ANALISYS

First of all, I neglected skin effect and try to found limits $\sigma \to 0$ and $\sigma \to +\infty$ (or $Z_2 \to \infty$ and $Z_2 \to 0$).

- In case $\sigma \to 0$ ($Z_2 \to \infty$) I found:

$\lambda_1 \to \infty$, $\lambda_2 \to 0$, $\alpha\beta \to \dfrac{-Z_{cs}(N-1)}{Z_3}$, $\chi \to 2(N-1)$, $\alpha\gamma \to -\dfrac{Z_1(N-1)+Z_{cs}}{Z_1 Z_3}$. And, therefore,

$$Z \to \dfrac{-\dfrac{Z_{cs}(N-1)}{Z_3} - 2(N-1)}{-\dfrac{Z_{cs}+Z_1(N-1)}{Z_3 Z_1} - \dfrac{2}{Z_1}}.$$ The last relation can be rearranged to

$$Z \to \dfrac{Z_1(N-1)\cdot(2Z_3 + Z_{cs})}{Z_1(N-1)+(2Z_3 + Z_{cs})}.$$

It corresponds to a circuit like impedance $Z_1(N-1)$ connected in parallel to impedance $(2Z_3 + Z_{cs})$. As usually $|Z_1(N-1)| \ll |2Z_3 + Z_{cs}|$ and there is no resonance, then $Z \to Z_1(N-1)$. It is just impedance of the empty cell!

- In case $\sigma \to \infty$ ($Z_2 \to 0$) I found:

$\alpha \to \dfrac{Z_1}{Z_3}$, $\lambda_{1,2}$ do not depend on $\sigma$, consequently, $\left( \dfrac{(\lambda_2^{N-1} - \lambda_1^{N-1} + \lambda_2 - \lambda_1)}{(\lambda_2^N - \lambda_1^N)} - 1 \right)$ is a constant as well as $\alpha$.

$\chi$ has a component that is reciprocal to the second degree of $Z_2$, when $\alpha\beta$ - just to the first, therefore, $\alpha\beta$ can be neglected. $\alpha\gamma$ has a component that is reciprocal to the second degree of $Z_2$.

So $Z \to \dfrac{\chi}{\alpha\gamma} \cdot \left( \dfrac{(\lambda_2^{N-1} - \lambda_1^{N-1} + \lambda_2 - \lambda_1)}{(\lambda_2^N - \lambda_1^N)} - 1 \right)$. As $\dfrac{\chi}{\gamma} \to \dfrac{\dfrac{2Z_{cs}Z_1}{Z_2^2}}{-\dfrac{1}{Z_1}\cdot\dfrac{Z_{cs}Z_1}{Z_2^2}} = -2Z_1$, then

$$Z \to -2Z_3 \cdot \left( \dfrac{(\lambda_2^{N-1} - \lambda_1^{N-1} + \lambda_2 - \lambda_1)}{(\lambda_2^N - \lambda_1^N)} - 1 \right).$$

This relation takes into account but does not include $Z_{cs}$!

Finally, if in addition $\left|\sqrt{\alpha}\right| \ll 1$ and $N \gg 1$, then

$$\left(\frac{\left(\lambda_2^{N-1} - \lambda_1^{N-1} + \lambda_2 - \lambda_1\right)}{\left(\lambda_2^N - \lambda_1^N\right)} - 1\right) \approx -\frac{\alpha}{2}(N-1) + \frac{\alpha^2}{24}N(N^2-1)$$

and, consequently, $Z \to Z_1(N-1)\left(1 - \frac{Z_1}{Z_3} \cdot \frac{N^2}{12}\right)$. As $\left|\sqrt{\alpha}\right| \ll 1$ and $N \gg 1$, then

$$Z_1(N-1)\left(1 - \frac{Z_1}{Z_3} \cdot \frac{N^2}{12}\right) \approx Z_1(N-1) \cdot \frac{1}{1 + Z_1(N-1) \cdot \frac{N}{12Z_3}}.$$

Or

$$Z = \frac{Z_1(N-1) \cdot \frac{12Z_3}{N}}{\frac{12Z_3}{N} + Z_1(N-1)}.$$

By analogy with previous limit here is impedance of a circuit like impedance $Z_1(N-1)$ connected in parallel to impedance $\frac{12Z_3}{N}$.

Also it is reasonable to calculate one more useful parameter: difference between the limits $\Delta$. Considering nonresonance case of $|Z_1(N-1)| \ll |2Z_3 + Z_{cs}|$, $\left|\sqrt{\alpha}\right| \ll 1$ and $N \gg 1$

$$\Delta = \lim_{\sigma \to \infty} Z - \lim_{\sigma \to 0} Z \approx Z_1 N\left(1 - \frac{Z_1}{Z_3} \cdot \frac{N^2}{12} - \left(1 - \frac{Z_1 N}{(2Z_3 + Z_{cs})}\right)\right) = (Z_1 N)^2 \left(-\frac{1}{Z_3} \cdot \frac{N}{12} + \frac{1}{(2Z_3 + Z_{cs})}\right) =$$

$$= (Z_1 N)^2 \left(\frac{12Z_3 - N(2Z_3 + Z_{cs})}{12Z_3(2Z_3 + Z_{cs})}\right) = -\frac{(Z_1 N)^2 (Z_3(2N-12) + NZ_{cs})}{12Z_3(2Z_3 + Z_{cs})} \approx -\frac{Z_1^2 N^3 (2Z_3 + Z_{cs})}{12Z_3(2Z_3 + Z_{cs})} =$$

$$= -\frac{Z_1^2 N^3}{12Z_3} = -(Z_1 N)^2 \cdot \frac{N}{12Z_3}.$$

Unfortunately I failed when trying to find a simple way to estimate sample conductance range corresponding to maximal sensitivity, so one have to use general formulas deduced above.

The next step is taking into account skin effect. In that case I used formula for $Z$ without any approximations or reductions. Values of parameters are given below (I did not know exact value $\varepsilon_1$ for the chemical test tube that I used to made the cell, so I chose just mean value for Pyrex):

$\varepsilon_S = 80$, $\mu = 1$, $\varepsilon_1 = 7$, $\varepsilon_2 = 1$, $R = 7.25\ mm$, $l = 12.5\ cm$, $d = 0.67\ mm$, $\delta = 20\ \mu m$, $h = 1.1\ mm$, $N = 155$, $L_0 = 40\ \mu H$, $R_0 = 0.9\ \Omega$.

The limit at high specific conductance (more than about 100 $S/m$) allow to split the model into two parts. The first model is valid for low-conductive samples (less than about 100 $S/m$, aqueous solutions) and do not take into account skin effect. The second part is appropriate for high-conductive samples (metals) and do not take into account coil-sample capacitance.

The result of calculation is presented at Fig. 3 (here and below skin effect is taken into account). Conductivity of a sample has a crucial effect on real part of $Z$, value can increase in more than 1000 times! It reflects ohmic losses in the sample. The second peak at high values of conductivity is due to eddy currents. The first peak represents influence of coil-sample capacitance. Real and imaginary parts are decreasing at high conductivity due to displacement of magnetic and, therefore, induced electric field to lateral surface of the sample. It results in decreasing of inductance and accessible to the current cross-section area of the sample. Increasing in current is not enough to compensate the decrease in the area, so ohmic losses decrease at high conductivities.

At the first blush, a conductivity meter with an inductive cell should be designed to measure real part or argument of cell impedance. But sensitivity drops rapidly with decreasing in frequency, while at high frequencies it is difficult to measure real and imaginary parts separately. Resonance circuits may be a solution, but such circuits are difficult to simulate with proper accuracy, they require more accurate adjustment after constructing. More convenient parameter is an absolute value of the impedance. Besides, it is usually monotonic at conductivities lower than 100 $S/m$ that correspond to aqueous solutions.

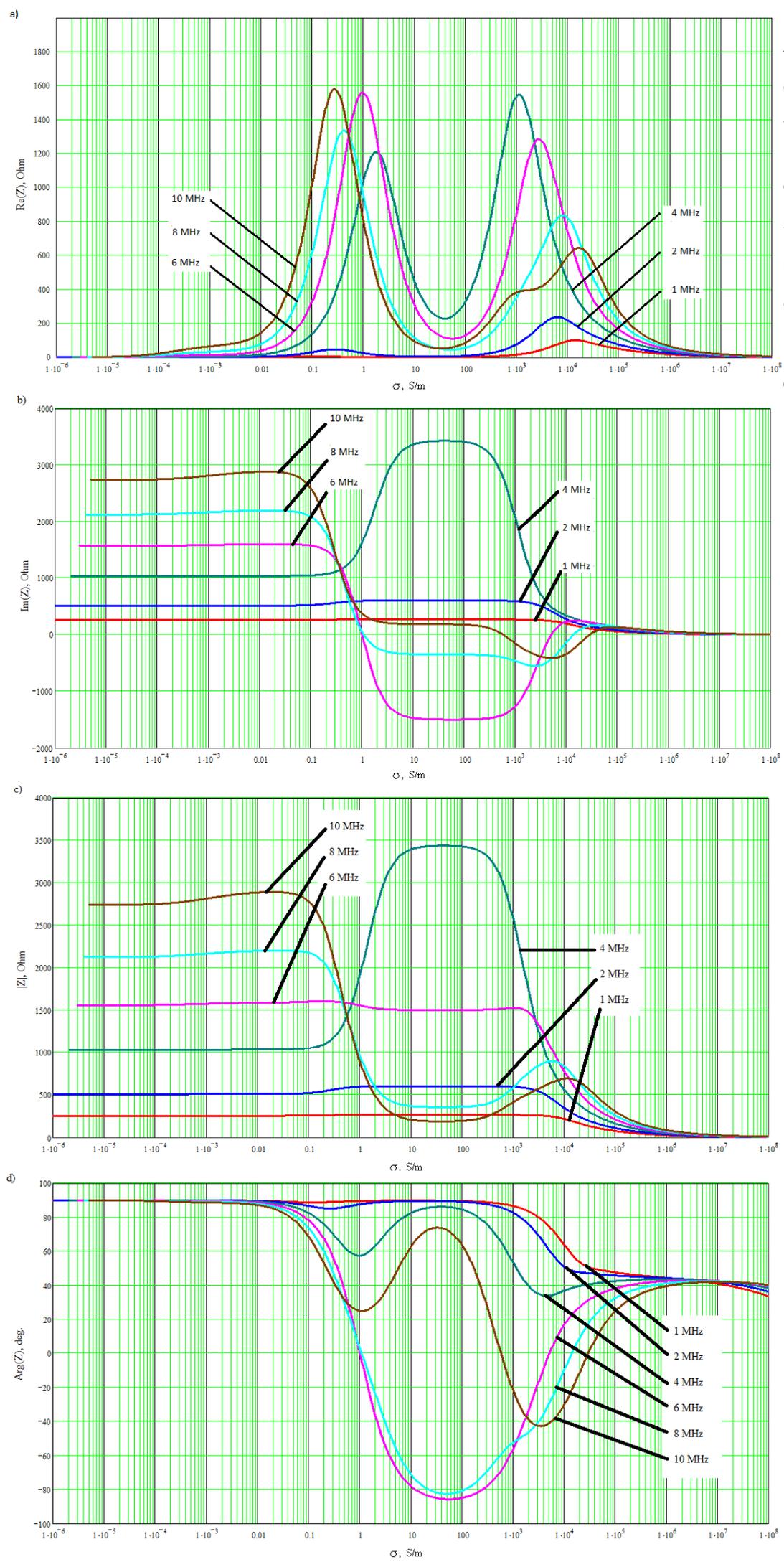

Fig. 3. The results of simulation: changes in impedance of the cell with respect to sample conductivity:

a) – real part of the impedance,
b) – imaginary part of the impedance,
c) – absolute value of the impedance,
d) – argument of the impedance.

EXPERIMENTAL TESTS

As there are a lot of satisfying models that are valid for high-conductive materials, the most interesting range of sample conductivity to test the model is from 0 to 100 $S/m$. Besides, I show above that it is possible to consider models for low- and high-conductive samples separately. So I used aqueous KCl or NaCl solutions of different concentration as samples and did not test the model for high-conductive substances like metals. Specific conductances of solutions with temperature correction were found with the help of reference book [6].

ABSOLUTE VALUE OF THE IMPEDANCE OF THE CELL

First of all, I tried to measure absolute value of $Z$ due to reasons mentioned above. I used LCR-meter that can operate at frequencies up to 1 MHz and has built-in averaging procedure and calibrating that takes into account parasitic capacitance of leads. Leads of the cell were short as far as possible and mutually spaced. So capacitance $C_W$ do not affect the results of measurement and I suppose $Z_{tr} \approx Z$. It is easy to compute that for empty cell at 1 MHz $|Z| \approx 251\,\Omega$ and difference between the limits $|\Delta| \approx 10\,\Omega$ (here I use parameters of the cell as presented above, it is characteristic of a real cell made from a standard chemical test tube and single-conductor copper wire). Experimental data are presented at Fig. 4b.

Firstly, I did not use measured $L_0$ and $R_0$, but I calculated it from geometric parameters of the cell and took into account skin effect in the wire for $R_0$ estimating. I did not know exact value of $\varepsilon_1$, so I used mean for Pyrex. Results of simulation are in qualitative accordance with the experimental study (Fig. 4a). When model $L_0$ and $C_2$ were adjusted (by varying $\varepsilon_1$ and winding density) to real (measured) parameters mentioned above, I obtained much more concordant results (Fig. 4b). It should be noted here that these two parameters have a crucial effect (in nonresonance case) and adjustment only them is enough. Moreover, $L_0$ is responsible for the limit at zero specific conductance and should be adjusted first, then $C_2$ adjustment gives limit at high conductance and at every intermediate point. By other words, only one high concentrated solution is enough for model calibrating.

IMPEDANCE OF THE CELL AT FREQUENCIES HIGHER THAN 1 MHz

Then I tried to measure absolute value of $Z$ at frequencies higher than 1 MHz. I did not have appropriate LCR-meter, so I used an oscilloscope and RF generator.

I used circuit presented at Fig. 5. Capacitor $C_d$ and the transducer (shown as impedance $Z$ connected to parasitic leads capacitance $C_W$ in parallel) form a voltage divider. Capacitor $C_u$ allow to neglect influence of the oscilloscope input capacitance (formed by circuits of the oscilloscope and coaxial cable of a test probe). Here I operated with another coil that had significantly high parasitic wire capacitance $C_W$, so I had to measure it firstly.

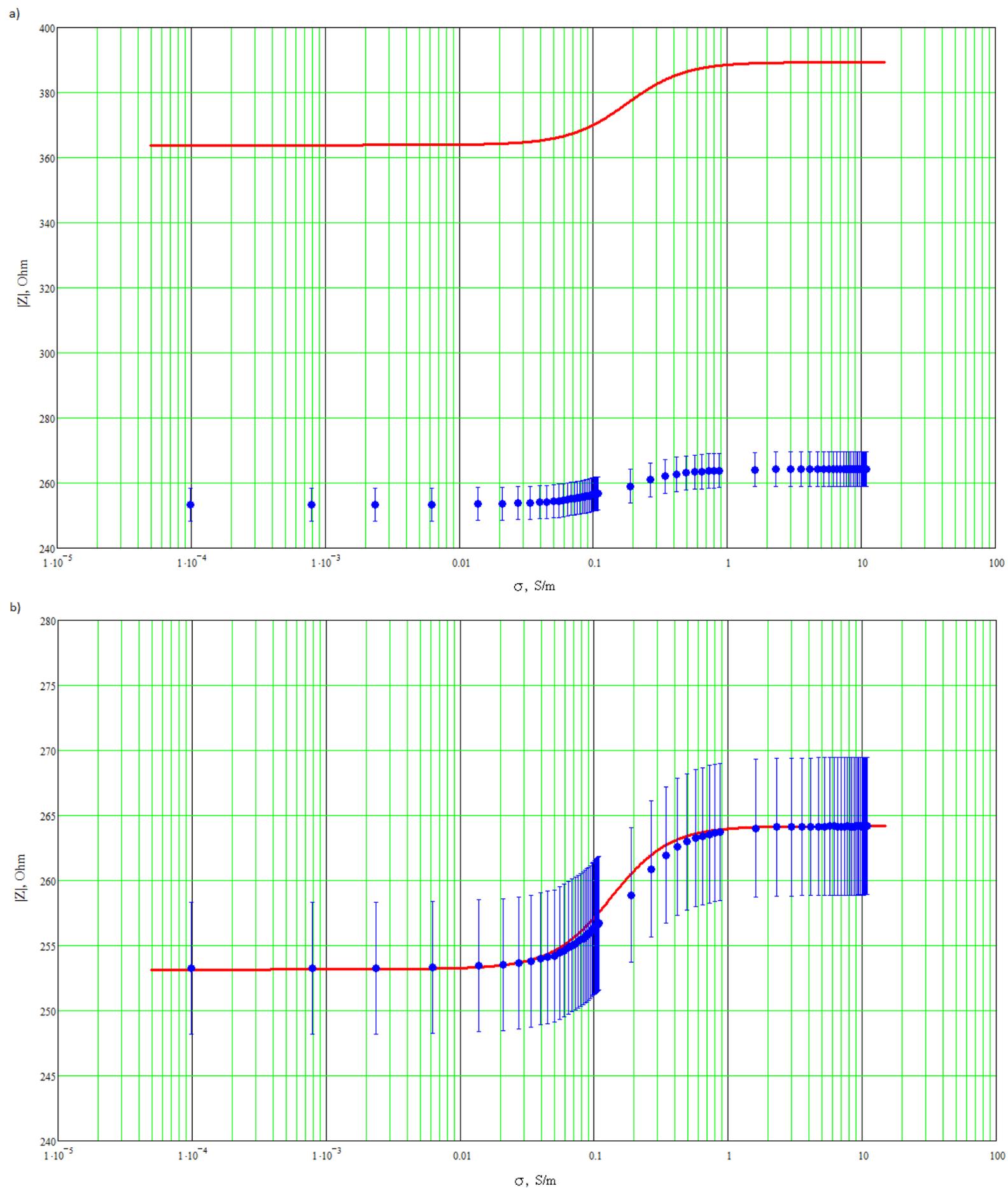

Fig. 4. Comparing experimental results and predicted by the model (at 1 MHz): a) – absolute value of the cell with respect to sample conductivity, the impedance has been calculated from dimensions of the cell and properties of materials; b) – absolute value of the cell with respect to sample conductivity, the impedance has been calculated from measured parameters of the empty cell ($L_0$ and $C_2$).

The measuring procedure is to determine frequency of the signal that corresponds to minimal amplitude at the output of the divider (while the cell is empty!, circuit at Fig. 5b). The frequency is resonance frequency for LC-circuit formed by the parasitic capacitance $C_W$ and inductance of the cell $L_0$. Knowing the frequency and inductance, $C_W$ is easily estimated. If, for any reason, the accuracy is not enough, one may determine the absolute value of transfer function (i.e. ratio of signal amplitudes at the output to the input of the divider) before each measurement with a solution while the cell is empty (Fig. 5a). Then, knowing the values of all the other elements of the circuit, $C_W$ can be calculated. However, it is more convenient to combine both methods and use following method. First, determine frequency range corresponding to the minimal signal amplitude (circuit at Fig. 5b). Second, choose value of $C_W$ inside the range and calculate transfer function. Finally, value of $C_W$ that gives no more than 5% difference between calculated and measured absolute values of the transfer function (while the cell is empty!, circuit at Fig. 5a) is target value.

Transfer function of the divider ($k$) can easily be calculated (circuit at Fig. 5a).

$$k = \frac{Z_{osc}}{(Z_{osc} + Z_{Cu})} \cdot \frac{(Z_{osc} + Z_{Cu}) Z_{tr}}{\left[(Z_{osc} + Z_{Cu}) Z_{tr} + Z_{Cd}(Z_{osc} + Z_{Cu} + Z_{tr})\right]},$$

where $Z_{Cu}$ and $Z_{Cd}$ are impedances of capacitors $C_u$ and $C_d$, respectively, $Z_{osc}$ - input impedance of the first oscilloscope channel (with coaxial cable of the test probe). It should be noted that due to high capacitance of the coaxial cable connecting divider output and oscilloscope there is difference between measured voltage and voltage at the divider output. This fact is taken into account by the first fraction of the formula.

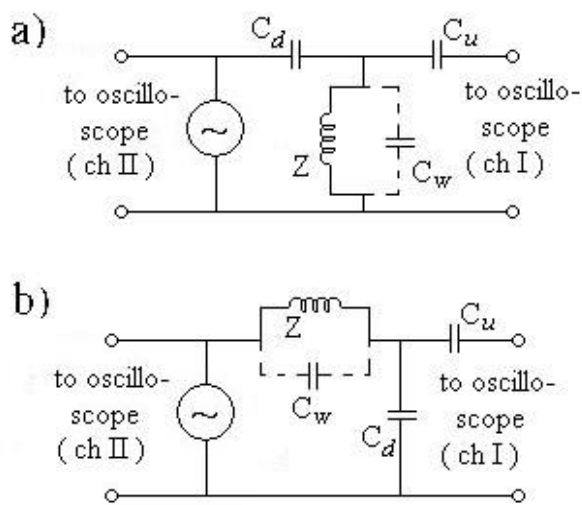

Fig. 5. Circuits of the experimental stand:

a) – variant to work with solutions,

b) – variant to measure parasitic cell leads capacitance $C_W$.

Knowing experimental values of $k$, $Z_{tr}$ can be calculated. But it is more convenient to compare experimental and simulated $k$. To test the model I measured transfer function ($k$, Fig. 5a) at different frequencies and concentrations of an electrolyte and compared it with given by the model. I used two similar transducers, the only difference is $C_2$ value (0.5 and 1.5 pF). Frequencies were chosen so that, by one hand, they were near resonance one. By other hand, I did not use frequencies that differed from resonance value less than 5% due to strict accuracy requirements. It allowed to combine significant amplitude difference with low accuracy of circuit parameters measurement. I used frequencies that differed from resonance one by 5-10%.

Frequency and capacitance $C_2$ affect transfer function dramatically. The response can be monotonic or not. It increase or decrease at different frequencies and capacitance. Some experimental data and results of simulation are given at Fig. 6.

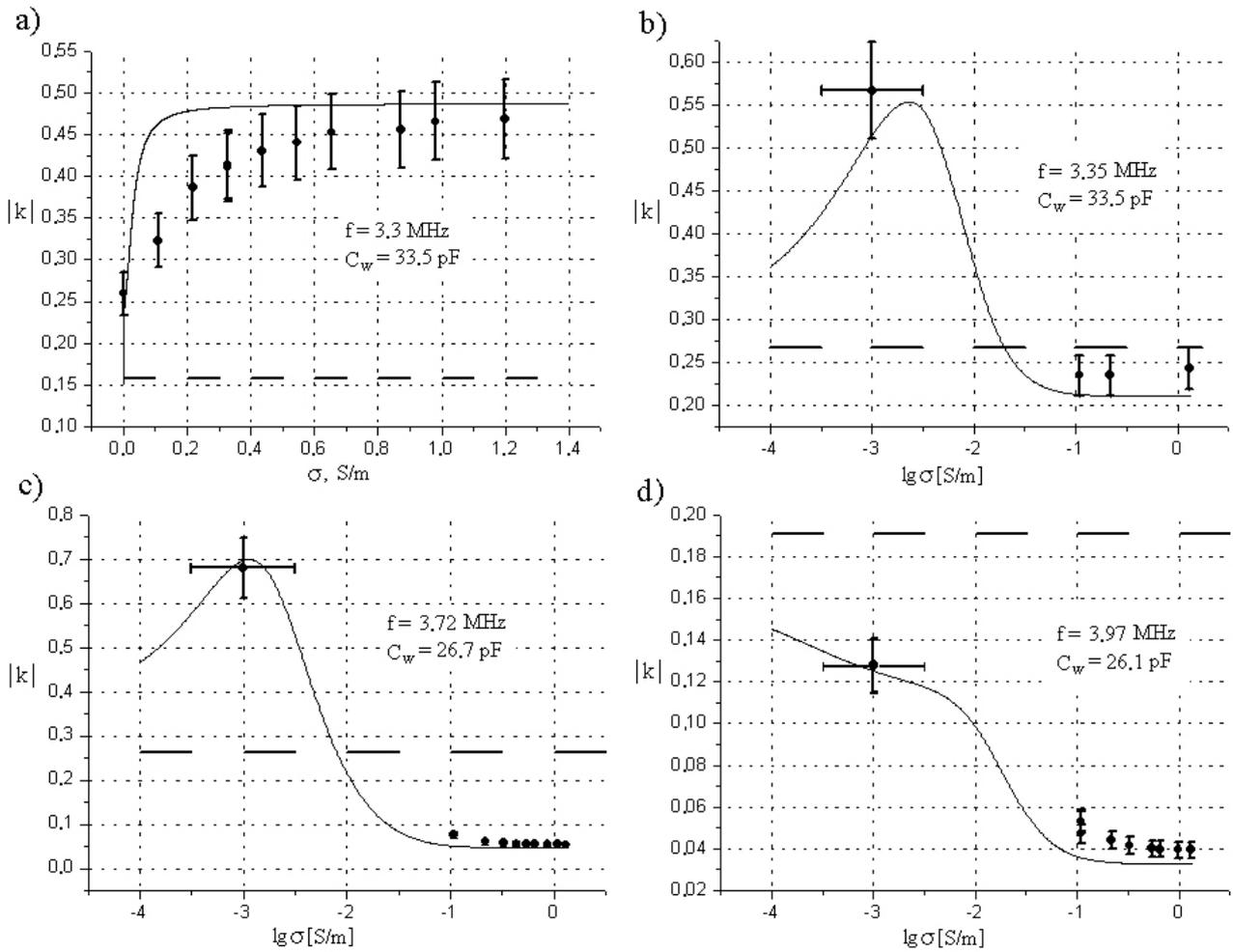

*Fig. 6. Simulated (solid line) and measured experimentally (points) dependences of the absolute value of the divider transfer function ( k ) on the electrical conductivity of the sample at various frequencies and parameters of the measuring cell (dotted line shows the experimentally obtained absolute transfer function value while the cell is empty).*

# OSCILLOMETRIC APPROACH TO THE CONDUCTOMETER DESING

Using a model to predict features of designed device is a good test for the model. So I tried to simulate behavior of a simple device.

As a first step to a conductometer, I studied two circuits of self-exciting oscillator presented at Fig. 7. The transducer ($Z_{tr}$) is a component of the oscillator, I used it as inductance. Changes in specific conductance of a sample were recorded as changes in frequency of oscillations. One of the circuits operate at series resonance (Fig. 7b), another – at parallel (Fig. 7a). Intermediate amplifier was set up as standard single-stage amplifier.

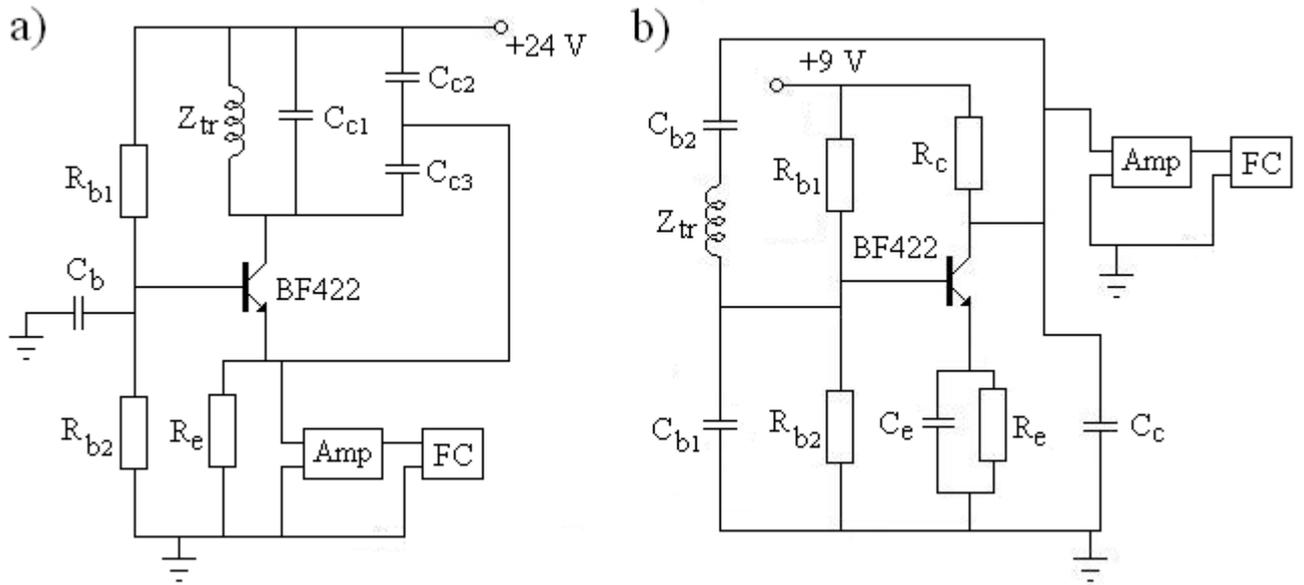

*Fig. 7. Circuits of self-excited oscillators: a) the oscillator operates at parallel resonance, b) – the oscillator operates at series resonance. Amp – intermediate amplifier, FC – frequency counter.*

It is know that self-exciting oscillator operates at frequency that causes to zero (or $2\pi$-fold) argument of loop gain. As parameters of a circuit is known, it is possible to plot $\varphi(\sigma,\omega)$ surface in a $(\omega,\sigma,\varphi)$ space (everywhere below $\varphi$ is loop gain argument). Frequency of oscillations is determined by intersection of the surface and plane $\varphi=0$. Regarding circuits presented at Fig. 7, loop gain ($K\beta$) is given by:

$$K\beta = \frac{1}{(1+SZ_{C3})}\left(\frac{Z_{C3}}{Z_{tr}}+\frac{Z_{C3}}{Z_{C1}}+1\right)\left(1+SZ_{C3}+\frac{Z_{C3}}{R_e}+\frac{Z_{C3}}{Z_{C2}}\right)$$ - for the oscillator shown at Fig. 7a,

$$K\beta = \frac{1+SZ_1}{1+SZ_1-SZ_2}\left(1+\frac{Z_2}{Z_{Cc}}+\frac{Z_2}{R_C}\right)\left(1+\frac{Z_2}{Z_{Cb1}}+\frac{Z_2}{R_{b2}}+\frac{Z_2}{R_{b1}}\right)$$ - for the oscillator shown at Fig. 7b.

Where $Z_{C1}$, $Z_{C2}$, $Z_{C3}$, $Z_{Cc}$, $Z_{Cb1}$ are impedances of respective capacitors, $S$ is averaged over period transconductance of the transistor, $Z_1 = \dfrac{R_e}{1+i\omega C_e R_e}$, $Z_2 = \dfrac{1+i\omega C_{b2} Z_{tr}}{i\omega C_{b2}}$.

Here I supposed that a transistor is an immediate-action component (i.e. imaginary part of parameter $S$ is equal to zero). I determine $S$ as amplitude of the first collector current harmonic divided by amplitude of alternating (sine) component of base-emitter voltage, $S$ was measured at 1 MHz. Determined in such manner parameter $S$ can significantly differ from transconductance measured at quasi constant current and depends on input signal amplitude (base-emitter voltage). Base current was neglected. Concerning the device presented at Fig. 7a, I supposed that base potential is a constant due to response time of base circuit more than 5 times bigger than period of oscillations.

Formulas concerning loop gain do not take into account parasitic inductance and capacitance of electronic components as well as nonlinear properties of transistors and anharmonicity of output signal oscillations. Every of these drawbacks can be responsible for significant difference between simulated and obtained frequencies of self-excited oscillations, even if there is no sample inside the cell. So I checked concordance of measured and predicted by the formulas frequencies, while the cell was empty.

For the oscillator operating at parallel resonance the difference was negligible (less than 0.25 MHz), but for other device the difference was about 1.2 MHz. It is worth to note that as higher frequency then more appreciable response of the cell to changing in sample conductivity (Fig. 3). Therefore, if there is significant difference between simulated and measured initial (with no sample in the cell) frequencies, one should expect only qualitative agreement of simulated and experimental data, while magnitude of frequency changes can vary greatly. However, this fact is not an obstacle for the use of a mathematical model of inductive cells. It reflects only drawbacks of formulas which were used to calculate loop gain.

Usually cell impedance response to changes in conductivity of a sample is a small correction to the impedance. Moreover, this small correction is independent of the parasitic parameters of the electronic components or transistor nonlinearity. Therefore, if one ensure agreement of calculated and measured initial frequencies (by manual adjusting frequency of the generator or by changing in values of parameters which determine frequency), one can expect not only qualitative but also quantitative agreement of calculated and experimental frequency-conductivity curves.

Considering the device operating at serial resonance, frequency-determining elements are the measuring cell and capacitor $C_{b1}$, parameters of the capacitor are independent of solution properties. So instead of the real value of capacitance $C_{b1}$ an effective one, that provides difference of initial frequencies not more than 0.25 MHz, can be used during calculating. And then effective value should be used in the calculation of the frequency-conductance curve. Considering the device operating at series frequency, I used exact this approach. Regarding the device operating at parallel resonance, I used real values of parameters during the calculation (measured directly or declared by manufacturer).

Simulated and measured frequency-conductivity curves presented at Fig. 8.

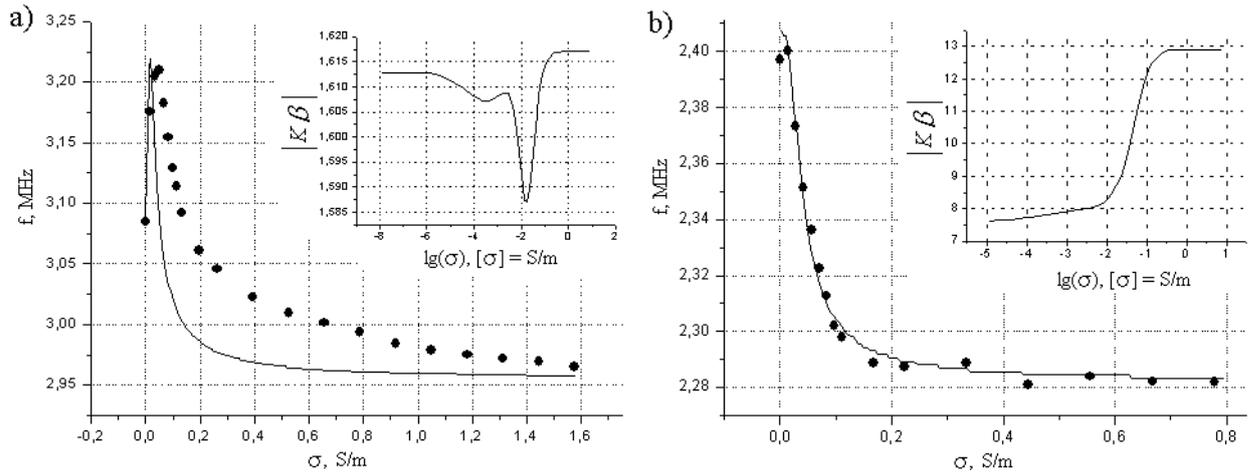

*Fig. 8. Calculated (solid line) and experimentally obtained (points) frequency of the oscillations with respect to the sample conductivity. Also shown calculated absolute loop gain values ($K\beta$): a) – for the device operating at parallel resonance, b) – for the device operating at series resonance.*

When designing an oscillometric conductivity meter, similar to those described above, one should pay special attention to the following fact. By increasing $C_2$ sensitivity increases sharply, especially in case of the low-conductive samples (0.01-0.1 $S/m$). But when the sensitivity increasing, there is unpleasant effect: if one take a cell that has relatively high $C_2$ value, quenching may occur at low sample specific conductance. This effect can be described in terms of the proposed equivalent circuit. Fig. 8 also shows absolute loop gain value (at $\varphi = 0$). It shows that while electrical conductivity of the sample is decreasing, considering the device operating at serial resonance, absolute loop gain value is also sharply decreasing. Regarding the device operating at parallel resonance, there is a local minimum. In this range of sample specific conductance quenching may occur. Sufficiently high values $C_2$ lead to disturbing of an amplitude balance. It can be shown that the greater the $C_2$ value the more significant decrease in absolute loop gain value.

As shown at Fig. 8 calculated absolute loop gain value is more than 1. It occurs due to *linear* approximation of the transistor parameters, for real transistor increase in input signal amplitude (base-emitter voltage) leads to decrease in parameter $S$, until absolute loop gain value reaches 1. Considering linear approximation, one should replace strict equation by inequation. Self-excited oscillations occur, while absolute loop gain value more than 1.

In conclusion I would like to note that, while designing oscillometric conductivity meter similar to described above, special attention should pay to following points.

- If one intends to operate the meter near sources of alternating electromagnetic fields, shielding of the entire device is required to improve reliability of measurements. However, the shielding may change the characteristics of the oscillator.

- Regarding an inductive transducer, to be able to neglect edge effects it is necessary to check that "long coil" condition is taking place. As for "short coil", the sensitivity may be higher, but that case is not described by a mathematical model proposed in this paper.

- To avoid the influence of frequency counter input impedance, it is desirable to use an intermediate amplifier that has a high input impedance and low input capacitance.

SUMMARY

Mathematical model of inductive measuring cell for contactless conductometry, that correctly describes the behavior of the inductive cell, is proposed in the article. The model takes into account not only effects associated with the coil-sample capacity ("capacitive effects"), but also influence of eddy currents induced into the sample under study and change in magnetic flux through the coil windings due to eddy currents. Equations of the model can be used to design conductometers. The model allows to estimate parameters of inductive measuring cell and helps to construct transducer that has desired properties.

ACKNOWLEDGMENTS

First and foremost, I would like to express my deep gratitude to Dmitry V. Knyazev, JIHT RAS laboratory of wide-range equation of state, for his friendship and active participation in this work. He provided a much-needed mathematical foundation that made this work possible.

I am grateful to Professor Eduard M. Trukhan, head of the MIPT subdepartment of biophysics and ecology, for five years of support and thoughtful guidance.

I would like to thank Alexander I. Dyachenko, head of GPI RAS laboratory of physics of living systems, for his help and invaluable support.

I acknowledge financial support from the Ministry of education and science of the Russian Federation. This work was supported by analytical departmental target program "Development of scientific potential of higher education" № 2.1.1/3179.

APPENDIX

When this text was ready for publication, another interesting effect had been observed. Changes of conductivity of a sample result in changes of amplitude-frequency characteristic of the transducer and, therefore, the divider transfer function is modified ($k$, Fig. 5a). If there is no sample, $|k|$ has only one maximum due to capacitance $C_d$ and inductance of the transducer. For diluted solutions with relatively low conductivity $|k|$ has also only one maximum (at another frequency). However, if there is concentrated solution (conductivity more than 1 S/m), pattern changes dramatically: other maxima arise (Fig. 9, Fig. 10).

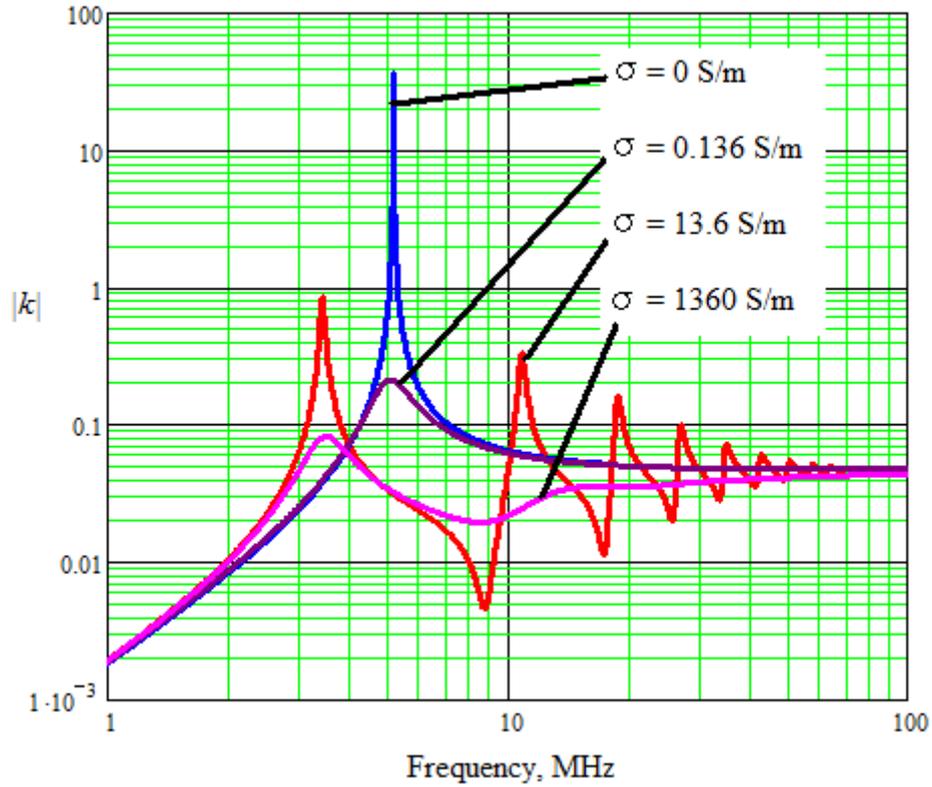

*Fig. 9. Simulated absolute values of the divider transfer function ($k$) for different samples.*

The effect was discovered by simulation and then confirmed experimentally. The maxima are vanishing, while conductivity is increasing. More likely, maxima reflect resonance properties of different LC-parts of intricate circuit presented at Fig. 2. The maxima arise, if influence of coil-sample capacitance becomes significant. In other words, if absolute value of impedance of the sample $Z_2$ approaches to or becomes less than the absolute value of $Z_3$.

For high-conductive samples the maxima are absent (conductivity more than 1000 S/m). As conductivity of a sample increases, influence of skin-effect rises, that leads to decrease of $Z_2$. At some conductivity absolute value of $Z_2$ becomes less than absolute value of $Z_3$. Situation is similar to taking place at low conductivities, the maxima vanish.

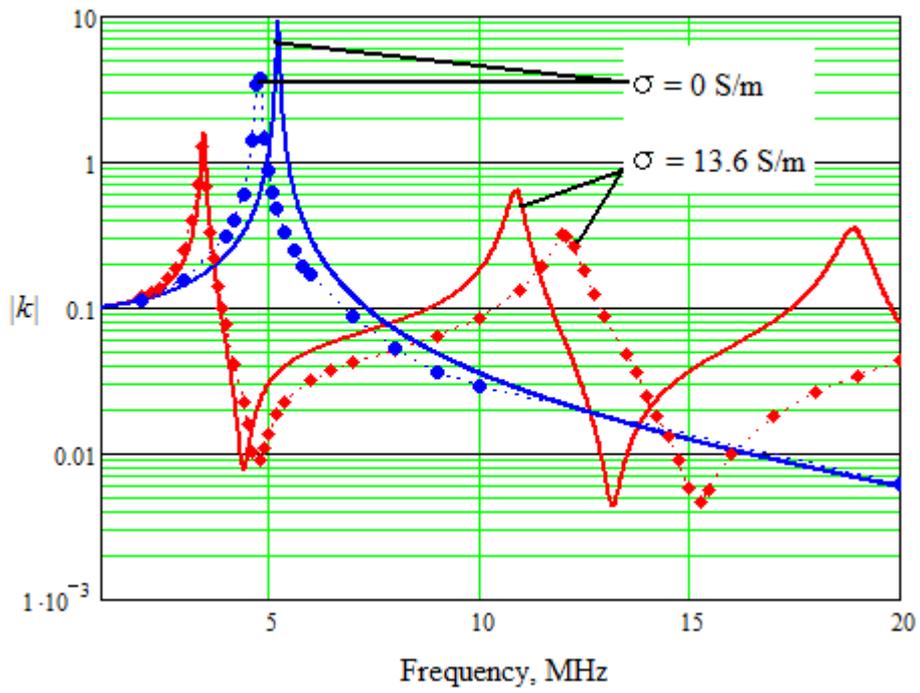

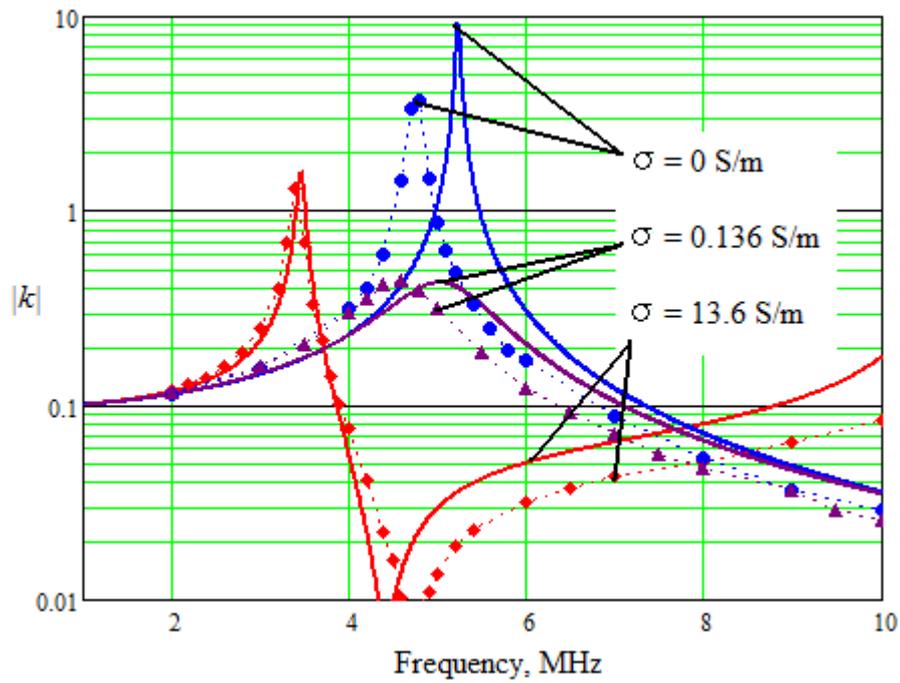

Fig. 10. Experimentally obtained (points) and simulated (solid) absolute values of the divider transfer function ($k$) for different samples.